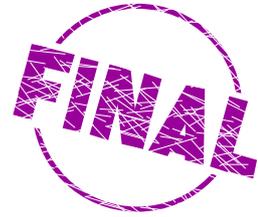

# Perturbation Theory with Unstable Fundamental Fields

Fyodor V. Tkachov

Institute for Nuclear Research of Russian Academy of Sciences, Moscow 117312, Russia

Monday, 16 November 1998, 12:56 pm

**Abstract**

The difficulties of perturbation theory associated with unstable fundamental fields (such as the lack of exact gauge invariance in each order) are cured if one constructs perturbative expansion directly for probabilities interpreted as distributions in kinematic variables. Such an expansion is made possible by the powerful method of non-Euclidean asymptotic operation [1].

Unstable particles ($Z$, $W$, top quark, higgs boson, etc.) are at the focus of current high energy physics research. However, a systematic theoretical formalism that would allow complete calculations for the processes involving such particles at the high level of precision required, say, for LEP2 [2], is still lacking (for recent discussions see [3], [4]). This Letter describes a systematic, inherently natural, and rather simple if unconventional modification of the standard perturbation theory (PT) that relieves the latter of the difficulties due to instability of heavy particles. (The physical situations we are concerned with are those where the issues of time evolution do not emerge; for a recent discussion of such issues see e.g. [5].)

Consider a process $q_1 \bar{q}_2 \to X \to l_1 \bar{l}_2$ mediated by an unstable field $X$ (e.g. a $Z$ or $W$ boson). Its amplitude $A(Q)$ contains the propagator of $X$ proportional to $[M^2 - Q^2 - i0]^{-1}$ where $M$ and $Q$ are the mass and 4-momentum of $X$ (for simplicity unitary gauge is implied throughout this Letter). The $-i0$ makes $A(Q)$ integrable around $Q^2 = M^2$ but the probability $P(Q) \propto |A(Q)|^2$ contains a non-integrable factor $[M^2 - Q^2]^{-2}$ which generates infinity when smeared over $Q^2$ around $Q^2 = M^2$. (Such a smearing is needed e.g. to take into account initial state QED radiation [6].) The standard PT expression is, therefore, meaningless.

Using an array of axiomatic techniques, Veltman [7] verified that a finite, unitary, and causal $S$-matrix in the Fock space spanned by stable particles only, is obtained if one Dyson-resums self-energy contributions $\Sigma(Q^2)$ corresponding to the instability ($g$ is the weak coupling):

$$\frac{1}{M^2 - Q^2 - i0} \to \Delta(Q; g) \equiv \frac{1}{M^2 - Q^2 - g^2 \Sigma(Q^2)}. \tag{1}$$

For unstable $X$, $\mathrm{Im}\,\Sigma(M^2) > 0$, so that the pole is pushed from the real axis into the complex plane and the corresponding probabilities are integrable.

Unfortunately, a well-known fact is that such resummations destroy exact gauge invariance (cf. e.g. [8]). This results in ambiguities of theoretical predictions — ambiguities which can be unsatisfactorily large [4]. Similarly spoiled is perturbative unitarity [9]. Somewhat similar ambiguities occur in QCD with renormalization scale fixing [10]. In QCD one usually limits the dimensional scale's variation to values around a typical dimensional scale of the process being studied. In the present case, however, the residual dependence on gauge-fixing parameters is completely unphysical and their variation cannot be limited from physical considerations. Various attempts to circumvent these difficulties [11], [12], [13], [14], [4] are ad hoc, unsystematic, and cumbersome. The issue can be traced back to the non-power dependence on $g$ in (1).

Yet an option for obtaining a well-defined systematic expansion in pure powers of $g$ does exist. To understand it, note the following:

(i) Although PT is usually developed for amplitudes, in the end one only needs probabilities $P(Q)$.

(ii) A *minimal physically motivated* restriction on mathematical nature of $P(Q)$ is that $P(Q)$ are measured using finite-resolution detectors and taking into account smearings for initial state radiation etc., so it suffices to define $P(Q)$ as distributions [15] for which only integration with smooth localized weights is defined but not necessarily pointwise values.

(iii) Even if exact $P(Q;g)$ are continuous in $Q$, perturbative expansion implies a limiting procedure $g \to 0$ which may (and actually does; cf. Eq. (2) below) bring one into the realm of singular distributions proper.

Now consider the probability (rather than the amplitude) of the same process $q_1 \bar{q}_2 \to X \to l_1 \bar{l}_2$ with the resummed $X$-propagator (1). Try to expand $|\Delta(Q;g)|^2$ in powers of $g$. The naïve Taylor expansion restores the usual PT expression — which is non-integrable and so is not a well-defined distribution. However, one actually deals with integrals of $P(Q)$ with arbitrary weights, and expansion in $g$ must preserve such integrability, i.e. the expansion must be *in the sense of distributions*. This means that one should expand not the product $|\Delta(Q;g)|^2$ per se but its integrals with arbitrary smooth weights. The whole point here is that it proves possible to obtain expansion formulas without specifying a concrete expression for the weight. That the expansion for such integral must contain something new compared with the naïve expansion of $|\Delta(Q;g)|^2$ is not surprising in view of the non-integrability of the naïve expansion.

A systematic theory of such expansions in the context of Feynman diagrams — the theory of asymptotic operation (AO) — has been fully developed[1] since 1982 [16] (for a review and complete references see [17]). Its Euclidean variant yielded powerful calculational formulas for short-distance and mass expansions [16], [18], [19], [20], [21] that are widely used at present [17], [22], [23]. The recent advance [1] extends AO to arbitrary problems in Minkowski space (with both loop and phase space integrals treated on equal footing).

The key discovery of the theory of AO is that distributions with respect to PT integrals play a role similar to that of complex numbers with respect to algebraic equations. In both cases the resulting freedom of constructive manipulation proves to be hugely useful (the psychological difficulties one encounters in both cases are also not dissimilar). Furthermore, the low awareness of the theoretical community at large of the distribution-theoretic foundation of the mentioned calculational formulas is rooted, on the objective side, in the fact that in Euclidean type problems, $\delta$-functions disappear from final formulas (similarly to how the imaginary unit may disappear from final expressions for the roots of algebraic equations). On the subjective side, the theory of AO emerged behind the Iron Curtain on the fringes of established theoretical communities. The two circumstances provided fertile soil for misinterpretations and miscitations to flourish (as discussed in [16] and [24]).

In contrast to Euclidean situations, the problem of unstable particles is the first application of AO where distributions cannot be eliminated from answers naturally. This is because the weak-coupling expansion in our case connects squared propagators which describe intermediate unstable fields and the $\delta$-functions which describe stable particles in the limit $g = 0$; cf. Eq. (2) and the discussion thereafter. So, the occurrence of singular distributions in a systematic expansion is rooted in the physical nature of the problem.

Our above example represents a simple exercise in application of the machinery of AO. Denote $\tau = M^2 - Q^2$, $h(\tau) = \mathrm{Re}\,\Sigma(Q^2)$, $f(\tau) = \mathrm{Im}\,\Sigma(Q^2)$, and $h_n, f_n = n$-th derivatives at $\tau = 0$. Then one has the following expansion for $g \to 0$,

$$|\Delta(Q;g)|^2 = g^{-2}\,\pi\delta(\tau)f_0^{-1} + \mathrm{VP}[\tau^{-2}] + \pi\delta(x)\left(h_1 f_0 - h_0 f_1\right)f_0^{-2} - \pi\delta'(x)h_0 f_0^{-1} + O(g^2), \qquad (2)$$

where the elementary distribution $\mathrm{VP}[\tau^{-2}]$ is defined by

$$\mathrm{VP}[\tau^{-n}] \equiv \frac{(-)^{n-1}\mathrm{d}^n \ln|\tau|}{(n-1)!\,\mathrm{d}\,\tau^n} \qquad (3)$$

with derivatives in the sense of distributions. The latter simply means that after integration with a smooth weight, the derivatives should be switched to the weight via formal integration by parts (see the excellent undergraduate level textbook [15].)

---

[1] Here "fully" means that, so far as I can see, all the formulas needed to apply the general theory of AO to expansions of concrete diagrams have been published. Of course, a big theory like this (the already published non-overlapping non-tutorial texts run well beyond 300 pages) can never be complete in the sense that its potential range of applications spans the entire theoretical particle physics based on Feynman diagrams ("No small parameter, no physics" [L.D. Landau]) so that there are many applications, each requiring a concretization of general formulas, as well as many topics of interest for mathematical physicists like finer points of proofs for various intermediate regularizations where, say, the dimensional regularization fails, generalizations to integrands beyond what may be ever needed in particle physics applications, etc.

To verify the expansion (2) [and its next term, Eq.(5)] is an elementary exercise in programming: choose any expressions for $f(\tau)$ and for $h(\tau) > 0$ [both analytical near $\tau = 0$], integrate both sides of (2) with any simple smooth weight over a finite segment containing $\tau = 0$, and vary $g$ to check how the error estimate scales with $g \to 0$.

If one recalls the omitted trivial factors and that $g^2 f_0 = M\Gamma_X$ to lowest order, then the $O(g^{-2})$ term in (2) corresponds to a free $X$-boson in the final state and to the familiar approximation

$$\sigma(q_1\bar{q}_2 \to X \to l_1\bar{l}_2) \approx \sigma(q_1\bar{q}_2 \to X) \times Br(X \to l_1\bar{l}_2). \tag{4}$$

The other terms on the r.h.s. of (2) yield a well-defined $O(g^2)$ correction to (4). The correction exactly coincides with $\tau^{-2}$ of the naïve PT for $\tau \neq 0$. The VP-prescription renders $\tau^{-2}$ integrable near $\tau = 0$, whereas the $\delta$-functional terms ensure the estimate $O(g^2)$ after termwise integration of (2) with smooth localized weights. So our method, while yielding unconventional but essentially elementary formulas, is in perfect agreement with both physical intuition and the standard PT: *the latter only misses the correct structure of the answer exactly at the singular point* [the VP-prescription and the $\delta$-functional terms on the r.h.s. of (2)].

The expansion (2) can be extended to any order in $g$; e.g. the term $O(g^2)$ is (the overall factor $g^2$ is dropped):

$$2h(\tau)\,\mathrm{VP}[\tau^{-3}] + \tfrac{1}{2}\pi\delta''(\tau)f_0^{-1}\left(h_0^2 - f_0^2\right) - \pi\delta'(\tau)f_0^{-2}\left(2h_0 h_1 f_0 - h_0^2 f_1 - f_1 f_0^2\right)$$
$$+ \pi\delta(\tau)f_0^{-3}\left(-2h_0 h_1 f_1 f_0 + h_0 h_2 f_0^2 + h_0^2 f_1^2 - \tfrac{1}{2}h_0^2 f_2 f_0 + h_1^2 f_0^2 - \tfrac{1}{2}f_2 f_0^3\right). \tag{5}$$

This, again, differs from the naïve PT expression $2h(\tau)\tau^{-3}$ only exactly at $\tau = 0$.

In the absence of massless particles and away from thresholds, the formulas (2)–(5) and their extensions to higher orders constitute a complete formalism (see, however, remark 6 at the end of the Letter).

The generalization to models with massless particles is obtained as follows. For simplicity we assume to be away from non-zero thresholds (the general method of AO works near thresholds too with appropriate modifications [1]). Furthermore, we assume there are no complications due to $t$-channel singularities in the physical region of the sort treated in [25]; such complications are tangential to what we are after.

The starting point is the $S$-matrix constructed according to the prescriptions of [7], but instead of amplitudes, one now deals with probabilities: One considers the collection of all unitarity diagrams with only stable particles in the initial and final states with self-energies responsible for instabilities Dyson-resummed.[2] Then one enforces expansions of products of propagators and phase-space $\delta$-functions in the sense of distributions in $g$ that occur in denominators (prior to any integrations). Mathematically, this resembles expansions in small masses and is done using the method of AO [1]. The result for each diagram is guaranteed to run in powers (only integer powers occur in the present case) and logarithms of $g$, with no other dependences on $g$ in the final result (the so-called perfectness of expansions which is a characteristic feature of AO [17]; concerning cancellations of logarithms of $g$ see below).

So, all one needs is to concretize the general prescriptions of non-Euclidean AO as presented in [1] to our specific problem. (This necessarily has to be done in the language of AO; consult [1], [20], [17]. Note that AO commutes with multiplications by polynomials [20], so numerators can be ignored in the present discussion and the description below is valid for fields of arbitrary spins.) It is sufficient to identify the *singular subgraphs* to which correspond counterterms to be added to formal expansion, together with the associated *transverse coordinates* and *scalings* needed to do power counting and reduce the coefficients of counterterms to power-and-log form. The following description of singular subgraphs is valid for models containing stable and unstable massive fields coupled to massless gauge fields (abelian or not). The case without unstable fields corresponds to QED/QCD-type soft singularities whose mechanism of cancellation is well understood; their cancellation takes place prior to expansions in self-energies, and the prescriptions given below correctly take into account those soft singularities on a diagram-by-diagram basis. In practice, the purely QED/QCD-type soft singularities do not require any additional special handling in the context of instability.

---

[2] It suffices to resum one-loop contributions. Inclusion of higher-order corrections into resummation does not affect the final answers; see below.

Let $G$ be a unitarity Feynman diagram. Call a line (ordinary or cut) massive or massless depending on the type of the corresponding field. Massive lines are further subdivided into unstable and stable. A _singular subgraph_ $\Gamma$ is a collection of some (one or more) massive (unstable, and, perhaps, stable) lines and, perhaps, some (zero or more) massless lines of $G$; $\Gamma$ should also satisfy the following restrictions (a straightforward concretization of the general completeness condition of [1], sec. 2.5):

a) Massless lines of $\Gamma$ should form a complete IR subgraph in the Euclidean sense [20]; let $k_a$ be their independent momenta.

b) Massive lines of $\Gamma$ should form "chains" as follows: set $k_a = 0$ and re-evaluate the momenta of all lines of $G$ using momentum conservation; denote their new values as $Q_b$; then a chain consists of all massive lines of $\Gamma$ with the same $Q_b$ and the same Lagrangian mass $M_b$ with no lines with the same $Q_b$ and $M_b$ in $G$ outside $\Gamma$. (A chain of $\Gamma$ may be part of a larger chain of $\Gamma' \supset \Gamma$ because the set of momenta $k_a$ depends on $\Gamma$.) A chain is unstable or stable depending on the corresponding particle's type.

If $\Gamma$ can be split into kinematically independent parts then its counterterm is a product of counterterms for its parts. Each such part should again be a singular subgraph in its own right. So, if the counterterm for any such part is zero (cf. below) then the counterterm for the entire $\Gamma$ is also zero. In particular, in a non-factorizing subgraph $\Gamma$, any its chain should be kinematically connected via massless lines to some other chain of $\Gamma$, meaning that there is a $k_a$ which passes through some (but not all) lines of either chain.

The following optional restrictions eliminate non-factorizable $\Gamma$'s whose counterterms are zero (as checked e.g. by explicit calculations):

i) Each unstable chain of $\Gamma$ should have a line on each side of the cut (otherwise the singularity is non-pinched and no counterterm is required).

ii) Similarly, each stable chain's $Q_b$ should correspond to an on-shell final/initial state particle; one can also say that each stable chain should contain a cut line (final or initial state).

The _transverse coordinates_ for such $\Gamma$ consist of all its $k_a$ and $\tau_b \equiv Q_b^2 - M_b^2$ for all its chains. Then the $\Gamma$-related counterterms to be added to the formal expansion are $\propto \prod_a \delta^{(D)}(k_a) \times \prod_b \delta(\tau_b)$ with total number of derivatives on all $\delta$-functions determined by power counting with _uniform scaling_ in all $\tau_b$ and $k_a$. This scaling also defines the homogenization needed to reduce counterterms to purely power-and-log form ([1], sec. 2.7).

The last important point concerns intermediate regularization of the formal expansion. Dimensional regularization alone is insufficient (as is clear from the above example). So one must introduce a separate VP-prescription for each "bad" product of unstable propagators as determined by simple inspection (cf. (2) and the example below).

With the above rules, writing out the AO for each unitarity diagram is a rather mechanical procedure ([1], Eqs. (2.9)–(2.12)) which yields the expansion in powers and logarithms of $g$. Explicit enumeration of singular configurations for physically interesting cases will be presented elsewhere [26].

A non-trivial example is given by the following diagram:

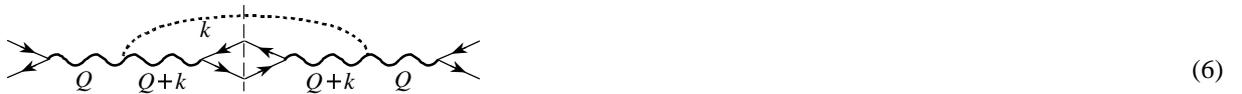

(6)

where $k$ corresponds to photon, $Q$ and $Q+k$, to a charged unstable $X$-boson. The corresponding product is

$$|\Delta(Q;g)|^2 |\Delta(Q+k;g)|^2 \delta_+(k^2), \tag{7}$$

where $\delta_+(k^2) = \theta(k_0)\delta(k^2)$. There are three singular subgraphs, each with one chain: $\Gamma_1$ and $\Gamma_2$ consisting of pairs of equal-momentum $X$-lines (the subproducts $|\Delta(Q;g)|^2$ and $|\Delta(Q+k;g)|^2$), and $\Gamma_3$ comprising the line $k$ and all $X$-lines. With $\tilde\tau = M^2 - (Q+k)^2$, the expansion of (7) in $g$ to $o(1)$ is

$$\mathrm{VP}[\tau^{-2}] \times \mathrm{VP}[\tilde\tau^{-2}] \delta_+(k^2) + E(\tau) \times \mathrm{VP}[\tilde\tau^{-2}] \delta_+(k^2) + \mathrm{VP}[\tau^{-2}] \times E(\tilde\tau) \delta_+(k^2)$$
$$+ \left\{ C_0 \delta(\tau)\delta(k) + C_1 \delta'(\tau)\delta(k) + C_1^\mu \delta(\tau)\partial_\mu \delta(k) \right\} + o(1), \tag{8}$$

where $E(\tau)$ comprises all the $\delta$-functional terms on the r.h.s. of (2). Eq. (8) strictly follows the general pattern of AO: The first summand is just the formal expansion with necessary VP-regularizations; the counterterms $E(\tau)$ and $E(\tilde{\tau})$ correspond to the singular subgraphs $\Gamma_1$ and $\Gamma_2$, and the counterterm in the last line, to $\Gamma_3$. Since the naïve formal expansion $\tau^{-2}\tilde{\tau}^{-2}\delta_+(k^2)$ is linearly divergent by power counting (based on the described uniform scaling in $\tau$, $\tilde{\tau}$, and $k$) at the singularity for $\Gamma_3$ (localized at $\tau = \tilde{\tau} = k = 0$), the last line contains only $\delta$-functions and their first-order derivatives. As is usual in the theory of AO, the coefficients $C_i$ depend non-trivially on $g$. For instance, the $O(g^{-2})$ contribution to $C_0$ is

$$\frac{\pi^2 M^{-2\varepsilon}}{4g^2 f_0 M^2}\left(\frac{1}{\varepsilon} - \gamma - \ln 4\pi - 2\ln\frac{g^2 f_0}{M^2} + O(\varepsilon)\right) + O(\ln g), \tag{9}$$

where $\varepsilon = \frac{1}{2}(4-D)$. $C_1^\mu$ and $C_1$ are similar (without the overall $g^{-2}$). For $C_1^\mu$, the bracket contains neither poles nor logarithms. The poles $\varepsilon^{-1}$ ensure integrability near $\tau = k = 0$ of the first two lines of (8). The finite parts ensure the asymptotic estimate.

The modified PT thus obtained contains all diagrams of the standard PT (with only stable particles in initial and final states) with VP-prescriptions for unstable particles' mass shell singularities, plus counterterms with some simple $\delta$-functions times unambiguously defined coefficients (cf. (2) and (8)).

**Discussion**

**1)** The modified and standard PT coincide off the mass shell of unstable particles, so perturbative unitarity is unaffected. For the same reason, loop integrals are not affected and their calculation can be performed as usual (whatever that means). The UV counterterms for loop integrals in the MS scheme are exactly the same in the modified PT as in the standard one (cf. the treatment of UV renormalization in the theory of AO [27]).

**2)** Evaluation of the non-trivial coefficients such as (9) has to be done only once for a finite number of different configurations of singular factors. (This work is currently in progress [26].)

**3)** The logarithms of $g$ in expansions of individual diagrams (cf. (9)) are completely analogous from mathematical point of view to logarithms of masses in Euclidean expansions [17], [28]. In the above example, contributions $O(g^{-2}\ln g)$ are cancelled by the two diagrams with virtual photon on one side of the cut (ensuring a physically correct limit $g \to 0$ after multiplication by decay vertex factors). This is a special case of a general mechanism based on unitarity of Veltman's $S$-matrix [7]: The unitarity relation $T^+ + T = T^+ T$ connects complete sums of unitarity diagrams (the r.h.s.) with amplitudes (the l.h.s.). However, the expansion of amplitudes in powers of $g$ is well-defined and contains no $\ln g$ terms, so such terms must cancel in the sum of all diagrams for the r.h.s. The physical origin of $\ln g$ contributions is the same as for the soft-photon singularities in QED, and their cancellation follows a similar pattern, too.

**4)** If Veltman's $S$-matrix is gauge invariant order-by-order within the corresponding precision, i.e. if the sum of diagrams through $O(g^n)$ is gauge invariant up to $O(g^{n+1})$ corrections, then the same is true for the modified PT. But the latter is a purely power-and-log expansion, so a gauge dependence of the sum of all terms proportional to $g^n$ could not have been cancelled by terms proportional to $g^N$ with $N > n$. So the modified PT must be *exactly* gauge invariant in each order.

**5)** A more general view on different variants of PT discussed here is as follows. Consider the integral Dyson equations for Green's functions of stable and unstable fields. Reconstruct $S$-matrix for stable particles only from the Green's functions as dictated by the first principles of QFT. The various versions of PT are then obtained via the following two tricks variously combined: (A) iterations of the Dyson equations starting from the free propagators and Lagrangian interaction vertices; (B) expansions of resulting diagrams to eliminate self-energies from denominators. Veltman's prescription then corresponds to an incomplete expansion in self-energies (rather than a resummation) — incomplete to avoid spoiling the qualitative structure of poles. The naïve PT is then restored from Veltman's amplitudes via expansion in the residual self-energies in denominators (which operation does not commute with the squaring of amplitudes in the case of unstable particles). The modified PT bypasses this step and completes the expansion directly for probabilities.

**6)** A potentially confusing feature of the modified PT that has to be clearly understood, is that its predictions are singular if integrable functions (cf. (2)), and their convergence to the continuous exact answer (i.e. to the familiar Breit-Wigner type shape) for $|Q^2 - M^2| \lesssim M\Gamma_X$ is not pointwise but the so-called _weak convergence_, i.e. convergence of integrals with smooth weights; such weights can be chosen arbitrarily and after integration with such a weight, the convergence is the usual convergence of real numbers. It is this failure of the naïve pointwise convergence that is routinely misinterpreted (by incorrect analogy with hadron resonances[3]) as indicative of a "fundamentally non-perturbative" nature of the problem whereas what one actually deals with is an unfamiliar — but in no way pathological or unusable; not even complex or difficult; just *unfamiliar* (which is enough to stir up passions though) — type of convergence.

First of all, from the viewpoint of sheer numerics, with rapidly varying functions (such as the Z peak without the initial and final state QED radiation), to rely on a pointwise convergence would be a notion very ill-conceived indeed. This is because negligible horizontal shifts of the function shape (such as due to higher-order corrections to the pole position) may induce misleadingly dramatic vertical shifts of the function values for a given value of the argument (an instability quite similar to what I discussed in connection with jet observables [29]; the material of sec. 15 of that work is particularly relevant for the present discussion). This means that it would be proper here to use one of the integral type convergences. For instance, to use the $L_2$ convergence (least squares) is a step in the right direction (although one usually keeps the pointwise convergence at the back of one's mind even when actually using least squares in the end). But the $L_2$ convergence does not take one far enough because it is unnecessarily restrictive in regard of the objects it can handle. In our case, the limit of zero coupling produces a phase space $\delta$-function in the leading order in place of a continuous curve, and our modified PT yields a systematic expansion around that $\delta$-function. So the logic of the problem compels one to pass the $L_2$ convergence and go directly to the weak convergence if one wants to keep all options open.[4]

Furthermore, the fixation on the Breit-Wigner shape appears to be artificial in the light of the fact that the modified PT yields an expansion around the phase-space $\delta$-function corresponding to the stable particle at zero coupling (a key physical element of the formalism). A parametrization of cross sections prior to incorporating QED/QCD radiation in terms of VP's and $\delta$'s (cf. Eq. (2)), while being _completely equivalent to the conventional Breit-Wigner type parametrization as far as the numerical information both contain is concerned_, is also a much simpler option (remember its exact gauge invariance). This simplicity is in a stark contrast with the convoluted[5] formalisms collectively erected around the Breit-Wigner shape and currently used for fitting the Z peak.

Furthermore, the Breit-Wigner shape, however classical,[6] is only a means to an end — in our case, tests and the fitting of parameters of the Standard Model — and *the weak convergence is fully sufficient for that*. For instance, for the case of LEP1 where one deals with just five fundamental free parameters [30], one could agree on a few weights localized around the Z peak, giving preference to those for which experimental errors are smaller and convergence of PT expansions is faster (both conditions favour slower-varying weights).

From a different perspective, the approximation of a continuous curve by singular distributions is, in point of fact, hardly more counterintuitive than representing arbitrary functions on the real axis via linear combinations of periodic functions in the theory of Fourier transform where one also encounters difficulties with the naïve pointwise convergence to exact results (the well-known Gibbs phenomenon [31]). The difficulties are resolved by modifying the "raw" Fourier approximations, e.g. using the trick of Feier summation. Such tricks exploit a priori information about the exact answer (usually its regularity properties).[7] They constitute an established

---

[3] That one should be careful to avoid such analogies was impressed upon me by I.F. Ginzburg.

[4] The fitting criteria based on the $L_2$ convergence are actually special cases of those based on the weak convergence. Indeed, nothing prevents the weak convergence from being used with ordinary functions in place of $L_2$ convergence. Instead of comparing directly function values as in the $L_2$ case, one then compares weighted integrals of the function, and one can also employ some kind of a least squares type minimization criterion for fits (other options such as robust statistics are also allowed). If one takes very narrow weights, the $L_2$ criteria are recovered. These arguments have a much wider significance than our concrete problem: whenever one deals with approximations of sharply varying functions, the approximation criteria based on weak convergence are preferable. This in fact is very much why weak convergences and distributions emerged and are widely used in the theory of partial differential equations.

[5] therefore, requiring too much hand work thus defeating an efficient automation which, for instance, is badly needed to tackle the $O(10^4)$ one-loop diagrams that contribute to the processes studied at LEP2.

[6] and so anonymous-referee-safe ☺.

[7] In other words, the input for obtaining the answer comprises, say, the number and values of Fourier coefficients *plus* the information about regularity properties of the answer. That such seemingly amorphous information can be effectively used

field of research in applied mathematics (cf. the classical treatise [32]).

Correspondingly, restoring a continuous function from singular approximations such as Eq. (2) is entirely feasible with knowledge of continuity of the answer. For instance, one can take a lesson from experimentalists who routinely apply a binning trick to finite samples of discrete events (either measured or obtained via Monte Carlo generators): such samples are, mathematically, nothing but sums of $\delta$-functions — singular distributions — whereas the resulting histograms are ordinary functions. The transition involves what has to be mathematically interpreted as integration of the raw sum of $\delta$-functions with rectangular weights (bins) spread along the real axis. (This is discussed in detail in sec. 15 of [29]; the ideology and much of the formalism described therein remains valid in the case of general distributions.) Such a processing never raises any eyebrows. A similar binning can be applied to theoretical answers such as Eq. (2); however, smooth rather than rectangular bins would have to be used here.

A manual insertion of finite width into unstable propagators is also a variant of a regularization procedure based on available a priori information (analyticity properties). However, it ought to be performed *after* gauge-invariant expressions are obtained (including loop corrections). This is similar to inverting a set of expansions (2) for some standard left hand sides in order to re-express the VP-distributions and $\delta$-functions in terms of a set of standard non-singular shapes.

Lastly, the singular distributions are generally smeared away by the convolutions necessary to take into account QED/QCD radiation (use the definition (3) and formal integration by parts). This removes most of the sting from this issue, and reduces the whole matter to one of technical convenience rather than principle.[8]

**7)** The described modification of PT is insensitive to inclusion of higher order self-energy corrections into the Dyson resummation: The additional dependences on $g$ (of the coefficients $f_i, h_i$ in the above examples) are analytical and as such allow safe Taylor expansion which in fact is automatically taken care of by the method of AO (via the homogenization). Together with uniqueness of power-and-log expansions [33], [17] this means that the modified PT per se involves no ambiguity whatsoever inasmuch as unstable fields are concerned. Therefore, and in view of the generality of the axiomatic results on which it is founded, the modified PT emerges as *the* perturbation theory for models with unstable fundamental fields.

*Acknowledgements.* I am indebted to I.F. Ginzburg for suggesting the problem and a crucial encouragement, to M.L. Nekrasov for pointing out singular configurations involving soft radiation from stable initial/final state particles, and to D.Yu. Bardin and G. Passarino for providing their notes on the theory behind TOPAZ0 and ZFITTER [30], the two programs used to analyze the LEP1 data around the *Z* peak. I also thank for discussions E.E. Boos, V.I. Borodulin, G.V. Jikia, Yu.F. Pirogov, L.D. Soloviov and N.A. Sveshnikov. This work originated in the lively atmosphere of the workshops of QFTHEP series (INP MSU, Moscow) which provided financial support as did the Theory Division of CERN where I benefited from discussions with D.Yu. Bardin, S. Dittmaier, V.A. Khoze, J. Papavassiliou and M. Testa.

---

in numerical applications, is only surprising at a first glance. Suppose the solution one seeks is a point on a plane, and one knows that the exact solution belongs to a line known a priori, and one insists (for whatever reasons) on approximations that also lie on that line. If one manages to obtain a "raw" approximate solution whose precision is known to be satisfactory but which happens to lie off the line, one simply improves the approximation by choosing the point on the line which is closest to the raw approximation; one then expects this new "regularized" approximation to approach the exact solution at the same rate as the raw one while staying on the line. The a priori information about, say, continuity of the exact solution is equivalent to stating that the latter lies in the subspace of continuous functions in the space of arbitrary functions. The infinite dimensionality of, and different convergences possible in, functional spaces creates specifics addressed by the theory of regularization methods. Such methods usually involve more or less sophisticated optimization techniques (finding an optimal point on the line closest to the raw approximation in the above example) to transform a raw approximation into a regularized one. The term "regularization" itself is motivated by the fact that such procedures are normally used to obtain approximations that possess required regularity properties starting from ones that do not.

[8] Unless one insists on "gedanken" experiments with monochromatic neutrino beams, in which case one *must* study the regularization methods mentioned above.